\documentclass[preprint,12pt]{elsarticle_new}
\usepackage{mathrsfs}
\usepackage{amsfonts}
\usepackage[all]{xy}
\usepackage{amssymb,bm}
\usepackage{amsmath}
\usepackage{amsthm}
\usepackage{bbding}
\usepackage{hyperref}
\newtheorem{thm}{Theorem}[section]

\newtheorem{cor}[thm]{Corollary}
\newtheorem{prop}[thm]{Proposition}

\newtheorem{conj}{Conjecture}

\newproof{pf}{Proof}
\newcommand{\qedd}{\hspace*{\fill}$\Box$\medskip}   %end box of a proof
\def\rk{\hbox{\rm{rank\,}}}
\def\det{\hbox{\rm{det}}}
\def\char{\hbox{\rm{Char}}}

\def\tr{\hbox{\rm{tr}}}

\def\Det{\hbox{\rm{Det}}}
\def\ker{\hbox{\rm{ker}}}

\begin{document}

\begin{frontmatter}

\title{Towards a conjecture on a special class of matrices over commutative rings of characteristic 2}

\author[1,2,3]{Baofeng Wu
\corref{cor1}
}
\ead{wubaofeng@iie.ac.cn}

\cortext[cor1]{Corresponding author}

\address[1]{Institute of Information Engineering, Chinese Academy of Sciences,\\ Beijing 100085, China}
\address[2]{State Key Laboratory of Cryptology, P. O. 
Box 5159,\\ Beijing 100878, China}
\address[3]{School of Cybersecurity, University of Chinese Academy of Sciences,\\ Beijing 100049, China}
\begin{abstract}
In this paper, we prove the conjecture posed by Keller and Rosemarin at Eurocrypt 2021 on the nullity of  a matrix polynomial of a block matrix with Hadamard type blocks over  commutative rings of characteristic 2. Therefore, it confirms the conjectural optimal bound on the dimension of invariant subspace of the Starkad cipher using the HADES design strategy. Moreover, we reveal  the algebraic structure formed by Hadamard matrices over commutative rings from the perspectives of group algebra and polynomial algebra. An interesting relation between block-Hadamard matrices and Hadamard-block matrices is obtained as well.
% This can further help us to generalize the definition of Hadamard matrices over commutative rings of any characteristic $p$.

\end{abstract}


\end{frontmatter}

%-------------------------------------------------------------------------------------------------------------------

\section{Introduction}\label{secintro}

At Eurocrypt 2021, Keller and Rosemarin posed the following conjecture in \citep{KR21} (an initial version appeared on \href{https://eprint.iacr.org/}{ePrint} in Feb. 2020 \footnote{See \href{https://eprint.iacr.org/eprint-bin/versions.pl?entry=2020/179}{https://eprint.iacr.org/eprint-bin/versions.pl?entry=2020/179}}), in their study of the resistance of the HADES design against invariant subspace attacks.

\begin{conj}[See {\citep[Conjecture 1]{KR21}} ]\label{conj1}
Let $k, s \in\mathbb{N}$. Let $R$ be a commutative ring with characteristic 2, and let $M$ be an $s\times s$ block matrix over $R$, each of whose blocks is a $2^k\times 2^k$ special matrix. Denote the blocks of $M$ by $\{M_{i,j}\}_{i,j=1}^s$. Let $M''\in R^{s\times s}$ be  defined by $M''_{i,j}=\lambda(M_{i,j})$, where $\lambda(M_{i,j})$ is the unique eigenvalue of the special matrix $M_{i,j}$. Denote by $q(x) = f_{M ''}(x)$ the characteristic polynomial of $M ''$.
Then $q(M)^2=0$.
\end{conj}

In Conjecture \ref{conj1}, a $2^k\times 2^k$ special matrix over a commutative ring\footnote{All rings considered in this paper are assumed to be unital ones.} $R$ is defined recursively in the manner that
$$M=
\begin{pmatrix}
    A&B\\
    B&A
\end{pmatrix},
$$
where $A$ and $B$ are both $2^{k-1}\times 2^{k-1}$  special matrices over $R$ (see \cite[Definition 1]{KR21}). Note that when $R=\mathbb{F}_{2^n}$ is a finite field, a special matrix is just the so-called Finite-Field-Hadamard (FFHadamard) matrix definied in \cite{sdm12}. Since when $\char(R)=2$ such special matrices share similar properties with the classical $\{\pm1\}$-valued Hadamard matrices,  in the following we also call them Hadamard matrices over $R$. See Section \ref{sec2:bghad} for some known properties of Hadamard matrices and their applications. 

\paragraph{Backgrounds on Conjecture \ref{conj1}} 
In recent years, one trend in the development of symmetric cryptography is to design specific symmetric ciphers for applications in advanced cryptographic protocals such as MPC, FHE and ZK. These ciphers are often known as arithmetization-oriented ciphers. Different from the design of classical symmetric ciphers,  the main goal in the design  of  arithmetization-oriented ciphers is to reduce the complexity of their arithmetic circuit implementations. Therefore, most of the known designs adapt different structures with classical structures such as Feistel and SPN. The HADES design strategy provides a good approach to design such new symmetric cipher structures.

The HADES design strategy \cite{hades}  combines the classical SPN structure with the partial-SPN (PSPN) structure \cite{pspn}. Recall that in a PSPN structure, the S-box is only applied to  a part of the state of each round. A HADES structure includes two layers of full SPN rounds at the head and tail of a cipher, and a middle layer of PSPN rounds. Obviously, this kind of design can reduce the number of S-boxes used in a cipher compared with classical SPN sturcture, and thus admits low arithmetic complexity. But on the other hand, since in the PSPN layer part of the input data remains unchanged through the S-box layer of each round, ciphers using the HADES structure may suffer from truncated differential attacks such as invariant subspace attacks. The designers will face new challenges in the design of linear diffusion layers of such ciphers, while in the classical design, MDS matrices are often enough.

In 2019, Grassi et al. \cite{sta} announced their designs of two families of  arithmetization-oriented hash functions, Starkad and Poesidon, aimed at applications in practical ZK proof systems. Both ciphers use the HADES structure in the design of their underlining permutations, while the difference lies in the input data supported: the permutation for Starkad supports inputs from a finite field $\mathbb{F}_{2^n}$ whereas in the case of Poesidon the finite field is $\mathbb{F}_{p}$ for odd prime $p$. In the design of their linear diffusion layers of the round functions, both ciphers use MDS matrices  of the Cauchy type, in the constructions of which $2t$ pairwise distinct elements $\{x_i,~y_i,~1\leq i\leq t\}$ from the underling finite fields are needed. Specifically, for the Cauchy matrix used in Starkad, $y_i = x_i + r$ for any $1\leq i\leq t$ where $r$ is a chosen constant. Another special structure of their round functions is that, only one S-box is used in each round of the PSPN layer. This may admit infinite long invariant subspace trials for the PSPN layer, and further result in successful attacks on the ciphers.

An invariant subspace trial of the PSPN layer means a collection of differential trials in which the S-box in each round will never be active. Assume the matrix in the linear diffusion layer of the round function is $M$. We consider  
\[U=\left\{ x\in \mathbb{F}_{q}^t \mid (M^\ell\cdot x)_1=0~\text{for~any}~\ell\geq0\right\},
  \]
where $\mathbb{F}_{q}$ is the underling field and $(x)_1$ represents the first component of $x\in\mathbb{F}_{q}^t$. Note that $U$ is a vector subspace of $\mathbb{F}_{q}^t$ and in the cases of 
Starkad and Poesidon, it promises all the infinite long invariant subspace trials. We call $U$ the invariant subspace of the corresponding cipher. If $U$ has a high dimension, one may perform attacks on the ciphers combining some other techniques such as algebraic attacks. It is easy to derive $\dim U\geq t-\ell$ if $\ell$ satisfies $M^\ell=\sum_{i=0}^{\ell-1}c_iM^i$ for some $c_i\in\mathbb{F}_{q},~0\leq i\leq\ell-1$. Therefore, to get the lower bound of $\dim U$, one needs to compute degree of the minimal polynomial of $M$. This is easy for an instance of $M$, but may be quite difficult for a generic construction of $M$ such as the classes of Cauchy matrices used in Starkad and Poesidon. 

Keller and Rosemarin \cite{KR21}  performed a generic successful attack on the Starkad cipher by proving 
\begin{equation}\label{krbound}
\dim U\geq t-(k+1)s
\end{equation}
 for any $t\times t$ Cauchy-type MDS matrix $M$ used in Starkad, where $t=2^ks$. 
It is a great observation in their attack  that this special construction of Cauchy matrix in Starkad can be viewed as an $s\times s$ block matrix with $2^k\times 2^k$ Hadamard-type blocks over $\mathbb{F}_{2^n}$. The bound \eqref{krbound} was obtained by proving that $q(M)^{k+1}=0$, where $q(x)$ is the characteristic polynomial of $M''$ as defined in Conjecture \ref{conj1}. 
As discussed before, dimension of $U$ depends on the smallest power $\ell$ such that $M^\ell$ can be represented as $\mathbb{F}_{2^n}$-linear combinations of lower powers of $M$. The nullity of $q(M)^{k+1}$ admits $\ell\leq (k+1)s$ since $\deg q(x)^{k+1}=(k+1)s$. Therefore, to improve the lower bound for $\dim U$, one approach is to prove lower nullity degree of $q(M)$ such as $q(M)^2=0$ which will improve the lower bound to $t-2s$. This was left as an open problem in \cite{KR21} and is where Conjecture \ref{conj1} comes from.

\paragraph{Our work}
In this paper, we give an affirmative answer to the open problem given in \cite{KR21} by proving Conjecture \ref{conj1}. It turns out that the main argument leads to the proof is incredibly simple, as long as we have found the key point. The proof relies heavily on the algebraic properties of Hadamard matrices over commutative rings of characteristic 2. Therefore, we further study the algebraic structure of the set formed by all such special matrices. We can characterize this algebraic structure by the tool of group algebra and further by multivariate polynomial residue ring.
These characterizations can help us to get simpler or even one-sentence proof for the main result $q(M)^{k+1}=0$ obtained in \cite{KR21}. We also give a relationship between block-Hadamard matrices (i.e., Hadamard-type block matrices or in other words, Hadamard matrices over a matrix ring) and Hadamard-block matrices (i.e., block matrices with Hadamard-type blocks or in other words, matrices over the ring of Hadamard matrices). This will admits a variant of the result implied by Conjecture \ref{conj1}.
%, and to generalize the construction of Hadamard matrices to commutative rings of odd characteristics.

We remark that  in the published version (see \cite{poe}) of \cite{sta}, the Starkad cipher is droped partially due to the attacks in \cite{KR21}. 
So the goal of this paper is not to improve attacks on Starkad. On one hand, Conjecture \ref{conj1} is indeed a theoretic problem arising from cryptanalysis of symmetric primitives, and a proof can fill the gap in the theory; on the other hand, by studying Conjecture \ref{conj1}, we can reveal deeper properties of Hadamard matrices, which should be help in future designs of symmetric ciphers applying such matrices.

\paragraph{Outline} 
The rest of the paper is arranged as follows. In Section \ref{sec2:bghad}, we recall some known properties of Hadamard matrices over commutative rings of characteristic 2. In Section \ref{sec3:proof} we give proof of Conjecture \ref{conj1} and some discussions will follow in Section \ref{sec4:discuss}. In Section \ref{sec:prophad} we characterize algebraic structure of the set of all Hadamard matrices. In Section \ref{sec:blockhad} we talk about the relationship between block-Hadamard matrices   and Hadamard-block matrices. 
% , which will help us to generalize them to commutative rings of odd characteristics in Section \ref{sec:hadgen}. 
Conclusions and further discussions will be given in Section \ref{sec:conclu}.

\section{Known properties of Hadamard matrices}\label{sec2:bghad}
%Let $R$ be a commutative ring with characteristic 2. 
Hadamard matrices over  a commutative ring $R$ have many nice properties. For example, the set of all $2^k\times 2^k$ Hadamard matrices, $\mathcal{H}_k(R)$, forms a commutative ring (see \cite[Proposition 1]{KR21}), and since it is naturally an $R$-module, it forms a commutative $R$-algebra. We further characterize structure of this algebra in Section \ref{sec:prophad}  to help understanding properties of Hadamard matrices deeper.

It is easy to observe that any $H\in \mathcal{H}_k(R)$ is determined by its first row, say, $(a_0, a_1, \ldots, a_{2^k-1})\in R^{2^k}$, from the recursive definition of a Hadamard matrix. By induction on $k$, one can prove that each element of $H$ can be determined by
\begin{equation}\label{hij}
    H_{i,j}=a_{i\oplus j},~0\leq i,j\leq 2^k-1.
\end{equation}
Note here that we index the rows and columns of $H$ starting from 0, and $\oplus$ is the exclusive-or operation of integers, in the sense of distinguishing them with binary vectors in $\mathbb{F}^k_2$ through 2-adic expansions. From this explicit representation of Hadamard matrices, one can derive all properties of them presented in \cite{KR21} in the case $\char(R)=2$, in a slightly different but more direct manner. We summarize some of them in the following proposition. 

\begin{prop}\label{prop1}
    Let $R$ be a commutative ring of characteristic 2 and $H, H_1, H_2 \in\mathcal{H}_k(R)$ where $k\in\mathbb{N}$. Let $\det(\cdot)$ and $\lambda(\cdot)$ denote the determinant and an eigenvalue of any matrix over a commutative ring. Then we have
    \begin{enumerate}[(1)]
        \item $H$ has a unique eigenvalue, namely, $\lambda(H)=\sum_{i=0}^{2^k-1}a_i$, where $(a_0, a_1, \ldots,$ $ a_{2^k-1})$ is the 1st row of $H$;
        \item $H^2=\lambda(H)^2 I_{2^k}$, where $I_{2^k}$ is the identity matrix;
        \item $\det(H_1+H_2)=\det(H_1)+\det(H_2)$;
        \item $\lambda(H_1+H_2)=\lambda(H_1)+\lambda(H_2)$, $\lambda(H_1H_2)=\lambda(H_1)\lambda(H_2)$.
    \end{enumerate}
\end{prop}

From an algebraic point of view, Proposition \ref{prop1} says that the two maps
$$\det :\;\mathcal{H}_k(R)\longrightarrow R ~\text{and} ~\lambda:\;\mathcal{H}_k(R)\longrightarrow R$$
are both homomorphisms of rings.

Let $R$ be a commutative ring with characteristic 2 and denote by $\mathscr{M}_{s\times s}(\mathcal{H}_k(R))$ and $\mathscr{M}_{s\times s}(R)$ the $\mathcal{H}_k(R)$- and $R$-algebra of $s\times s$ matrices over $\mathcal{H}_k(R)$ and $R$, respectively. The homomorphism $\lambda:\;\mathcal{H}_k(R)\longrightarrow R$ extends naturally to an $R$-algebraic homomorphism $$\bar{\lambda}:\;\mathscr{M}_{s\times s}(\mathcal{H}_k(R))\longrightarrow \mathscr{M}_{s\times s}(R),~\left(M_{i,j}\right)\longmapsto \left(\lambda(M_{i,j})\right) .$$
Similarly, the homomorphism $\det$ extends to $\overline{\det}$ over matrix algebras in this manner. For the purpose of clearity, let $\Det$ denote the classical determinalt map for $s\times s$ matrices over commutative rings, which is a  homomorphism between multiplicative monoids of rings.
Then Proposition \ref{prop1} also implies that the diagrams for multiplicative monoids of algebras (rings)
\[
    \xymatrix@C=2.4cm@R=1.5cm{
        \mathscr{M}_{s\times s}(\mathcal{H}_k(R))\ar[r]^{\Det}\ar[d]_{\bar{\lambda}} & \mathcal{H}_k(R)\ar[d]^{\lambda}\\
        \mathscr{M}_{s\times s}(R)\ar[r]_{\Det} & R
    }
  \]
and
\[
    \xymatrix@C=2.4cm@R=1.5cm{
        \mathscr{M}_{s\times s}(\mathcal{H}_k(R))\ar[r]^{\Det}\ar[d]_{\overline{\det}} & \mathcal{H}_k(R)\ar[d]^{\det}\\
        \mathscr{M}_{s\times s}(R)\ar[r]_{\Det} & R
    }
  \]
are both commutative (see also \cite[Proposition 8]{KR21}). 

In fact, when $R=\mathbb{F}_{2^n}$ is a finite field, properties of Hadmard matrices over $R$, namely, the FFHadmard matrices, appear in the literature before \cite{KR21}. They behave as a good source of involutory MDS
 matrices which are friendly in the design of linear diffusion layers for classical symmetric ciphers. We refer to \cite{you97,br00,sdm12,gup13} for some previous work related to them.
In addition, when $R=\mathbb{R}$ is the real field, we  recall another  interesting source of Hadmard matrices over $R$ in cryptography. Let $X$ be a random variable over $\mathbb{F}_{2}^k$ with probability distribution $(x_0, x_1, \ldots,x_{2^k-1})$, that is, $\Pr(X=i)=x_i$ for any $0\leq i\leq 2^k-1$ distinguished with a binary vector in $\mathbb{F}_{2}^k$ by 2-adic expansion. Let $X'$ be a random variable over $\mathbb{F}_{2}^k$ which is independent with $X$ with probability distribution $(x_0', x_1', \ldots,x_{2^k-1}')$. Then the probability distribution $(y_0, y_1, \ldots,y_{2^k-1})$ of the random variable $Y=X\oplus X'$ can be determined by
\[
(y_0, y_1, \ldots,y_{2^k-1}) = (x_0', x_1', \ldots,x_{2^k-1}')\cdot T_\oplus,
\]
where $T_\oplus$ is just the Hadamard matrix determined by the first row $(x_0, x_1, \ldots,$ $x_{2^k-1})$, namely, $(T_\oplus)_{i,j}=x_{i\oplus j}$, $0\leq i,j\leq 2^k-1$. This result is useful in truncated differential attacks of  symmetric ciphers; see \cite{hou23}.

\section{Proof of Conjecture \ref{conj1}}\label{sec3:proof}

In this part we explain how to prove Conjecture \ref{conj1}. It turns out that the main argument leads to the proof is very simple. 

For a generic $s\times s$ matrix $A=(a_{ij})$ over a commutative ring, it is clear from the definition of determinant  that $\Det(A)$ is a multivariate polynomial in the entries $\{a_{ij}\}$. Assume $f_A(x)=\det(xI_s-A)=x^s+\sum_{i=1}^sf_ix^{s-i}$ is the characteristic polynomial of $A$. We are then clear that $f_k$ is a multivariate polynomial in the entries $\{a_{ij}\}$ for any $1\leq k\leq s$. In fact, it is well known $f_s=(-1)^s\Det(A)$. A not-so-well-known result is that for $1\leq k\leq s$, 
$$f_k=(-1)^k\tr\left( {\bigwedge}^k A \right),$$
where $\tr( \wedge^k A )$ is the trace of the $k$-th exterior power of the endmorphism induced by $A$, which can be computed as the sum of all principle minors of $A$ of size $k$. Since each minor of $A$ is the determinant of a sub-matrix, of course a multivariate polynomial in the entries of $A$, hence  $f_k$ is also a multivariate polynomial in the entries of $A$ for any $1\leq k\leq s$.  

Let $R$ be a commutative ring with characteristic 2, and let $M$ and $M''$ be  the matrices in Conjecture \ref{conj1}. Instead of a block matrix, we view $M$ as a matrix over the commutative ring $\mathcal{H}_k(R)$. Assume the characteristic polynomial of $M$ and $M''$ are $Q(x)=\sum_{i=0}^sQ_ix^{i}$ and $q(x)=\sum_{i=0}^sq_ix^{i}$, respectively. Note that $Q_i\in \mathcal{H}_k(R)$ while $q_i\in R$, $0\leq i\leq s$. From the above discussion, $Q_i$ and $q_i$ can be computed by evaluating the same multivariate polynomial in the corresponding entries of $M$ and $M''$, respectively. Since $\lambda:\;\mathcal{H}_k(R)\longrightarrow R$ is a homomorphism, we are clear that $\lambda(Q_i)=q_i$. From Proposition \ref{prop1} (2), we also have $Q_i^2=\lambda(Q_i)^2\cdot id=q_i^2 \cdot id$, where $id$ is the identity element of $\mathcal{H}_k(R)$, namely, $I_{2^k}$.

By Cayley–Hamilton theorem for matrices over commutative rings, we know that $Q(M)=0$, and of course $Q(M)^2=0$. Since the ring  $\mathcal{H}_k(R)$ also has characteristic 2, we have
$$0=Q(M)^2=\sum_{i=0}^sQ_i^2M^{2i}=\sum_{i=0}^s(q_i^2\cdot id)M^{2i}=\sum_{i=0}^sq_i^2M^{2i}=q(M)^2.$$
This completes the proof.

\section{Further discussions on  Conjecture \ref{conj1}}\label{sec4:discuss}

The proof of Conjecture \ref{conj1} answers the second open problem in \cite{KR21}, that is,
the lower bound of dimension of the invariant subspace $ U$ defined for the $t\times t$ Cauchy-type MDS matrix $M$ used in the design of the Starkad cipher can be improved to $t-2s$ where $t=2^k\cdot s$.

A natural following question is whether this bound can be further improved.  We should note first that the bound $t-2s$ is a general one, not depending on the ring $R$ and the shapes of these Hadamard blocks of $M$. Of course when these blocks are of certain special types, e.g., scalar matrices, the bound $t-2s$ can be improved to, e.g., $t-s$. However, this is not the case for the Cauchy matrix used in Starkad. 

Another natural question is  whether the characteristic polynomial $q(x)$ in Conjecture \ref{conj1} can be replaced by minimal polynomial, which may has a degree less than $s$. More precisely, if the minimal polynomials of $M''$ is $\phi(x)$, shall we have $\phi(M)^2=0$?  First, it should be noted that in this case the method for proving Conjecture \ref{conj1} in Section \ref{sec3:proof} will not work, since coefficients of minimal polynomial of a matrix have no direct and explicit relations with its entries. Second, when $R$ is  a generic commutative ring, the minimal polynomial of a matrix over $R$ may not be unique. Actually, minimal polynomial of a matrix $A$ over $R$ is defined as the least degree polynomials in the annihilating ideal of $A$ in $R[x]$, which may not be a principle ideal. Even the minimal polynomial is unique, $\phi(M)^2=0$ does not always hold. 
Indeed, one can quickly observe that, for example, when $M''=0$, its minimal polynomial is $\phi(x)=x$, however, one cannot obtain $M^2=0$ for any $M$ whose blocks all have eigenvalue 0. 

But on the contrary, if we can find the minimal polynomial $\Phi(x)=\sum_{i=0}^s\Phi_ix^i\in\mathcal{H}_k(R)[x]$ of $M$, that is $\Phi(M)=0$, then we can obtain $\phi(M)^2=0$ where $\phi(x)=\sum_{i=0}^s\phi_ix^i$ with $\phi_i=\lambda(\Phi_i)$. 
This will improve the lower bound of $\dim U$ to $t-2\cdot\deg\Phi(x)$. However, for generic $s\times s$ matrices over commutative rings, the best general upper bound for the degrees of their minimal polynomials one can get is $s$. So in this sense, the bound $t-2s$ for a generic $M$ is optimal. When $M$ is considered in some special classes of  matrices over $\mathcal{H}_k(R)$, e.g., circulant matrices, Vandermonde matrices, or Hadamard matrices we consider, this bound can possibly be improved. 

As for Conjecture \ref{conj1}, it seems hard to directly prove it through evaluating $q(M)^2$. It has already been observed in \cite{KR21} that $q(M)$ lies in the kernel of the homomorphism $\bar{\lambda}$, that is, all blocks of $q(M)$ have eigenvalue 0. However, as mentioned above, we cannot obtain $\tilde{M}^2=0$ for any $\tilde{M}\in \ker\bar{\lambda}$ in general. Besides, we can see that if $q(M)^2=0$, then for any $\tilde{M}\in \ker\bar{\lambda}$, we have 
$$q(M+\tilde{M})^2=0.$$ 
But this does not mean if $g(M)=0$ for certain $g(x)\in R[x]$, then we have $g(M+\tilde{M})=0$ for any $\tilde{M}\in \ker\bar{\lambda}$. Indeed, any $M\in \mathscr{M}_{s\times s}(\mathcal{H}_k(R))$ can be factorized into 
$$
M=M''\otimes I_{2^k} + \tilde{M}
$$
for a unique $\tilde{M}\in \ker\bar{\lambda}$. Obviously, for any $g(x)\in R[x]$ with $g(M'')=0$ (e.g., $g(x)=q(x)$, the characteristic polynomial of $M''$), we have $$g(M''\otimes I_{2^k})=g(M'')\otimes I_{2^k}=0.$$ But this does not promise $g(M)=0$ for any $\tilde{M}\in \ker\bar{\lambda}$.

Another interesting corollary of the proved Conjecture \ref{conj1} is, if ${M}\in \ker\bar{\lambda}$, then we have $M^{2s}=0$ (not depending on $k$ which determines the size of each block) since the characteristic polynomial of $M''=0$ is $x^{s}$. Recall that in \cite{KR21} it was proved $M^{k+1}=0$, an equality depending on $k$. The power $k+1$ comes from \cite[Proposition 7]{KR21}, namely, any $k+1$ elements of $\mathcal{H}_k(R)$ all having eigenvalue 0 will multiply to 0. The result $M^{2s}=0$ implies more complicated relations between elements of $\mathcal{H}_k(R)$ having eigenvalue 0, which seems not easy to directly reveal. In the next section we will further discuss the set of all such elements.

\section{Structure of the algebra $\mathcal{H}_k(R)$}\label{sec:prophad}

To further understand properties of Hadamard matrices over a commutative ring $R$ (not necessarily with characteristic 2), in this part, we give characterizations of the structure of the algebra formed by them, namely, the $R$-algebra $\mathcal{H}_k(R)$.

Let $G=(\mathbb{F}_2^k,\oplus)$, the additive group of the vector space $\mathbb{F}_2^k$. We denote the identity of $G$ by $e$, i.e., $e=(0,0,\ldots,0)$. Let $R[G]$ be the group ring (algebra) generated by $G$ over $R$. Elements of $R[G]$ are all of the form $a=\sum_{g\in G}a_gg$ where $a_g\in R$ for any $g\in G$, that is, formal linear combinations of elements of $G$ over $R$. Multiplication of two elements $a$ and $b$ are defined in a convolutional manner, that is,
\begin{equation}\label{mulgpalg}
   \left(\sum_{g\in G}a_gg\right) \left(\sum_{g\in G}b_gg\right)=\sum_{g,h\in G}a_gb_h(g\oplus h)=\sum_{g\in G}\left(\sum_{h\in G}a_gb_{g\oplus h}\right)g. 
\end{equation}
We have the following theorem.

\begin{thm}\label{isogpring}
    \[\mathcal{H}_k(R)\cong R[G].\]
\end{thm}
\begin{pf}
For two Hadamard matrices $A$ and $B$ in $\mathcal{H}_k(R)$, assume their first rows are $(a_0, a_1, \ldots, a_{2^k-1})\in R^{2^k}$ and $(b_0, b_1, \ldots, b_{2^k-1})\in R^{2^k}$, respectively. Then we know from \eqref{hij} that 
\[A=\left(a_{i\oplus j}\right)_{i,j=0}^{2^k-1},~~B=\left(b_{i\oplus j}\right)_{i,j=0}^{2^k-1}. \]
Let $C=AB=(c_{ij})$. Then we have 
\[c_{ij}=\sum_{k=0}^{2^k-1}a_{ik}b_{kj}=\sum_{k=0}^{2^k-1}a_{i\oplus k}b_{k\oplus j}=\sum_{k=0}^{2^k-1}a_{ k}b_{k\oplus i\oplus j},\]
which means $C$ is a Hadamard matrix with first row $( \sum_{k=0}^{2^k-1}a_{ k}b_{k\oplus  j}\mid 0\leq j\leq 2^k-1)$. Therefore, the map 
$$\mathcal{H}_k(R)\longrightarrow R[G],~~(a_{i\oplus j})\longmapsto \sum_{j=0}^{2^k-1}a_{{\rm bin}(j)}{\rm bin}(j)$$
implies the isomorphism between $\mathcal{H}_k(R)$ and $R[G]$ according to \eqref{mulgpalg}, where
$${\rm bin}:~ \mathbb{Z}_{2^k}\longrightarrow G,~~j=\sum_{l=0}^{k-1}j_l2^{k-1-l}\longmapsto (j_0,j_1,\ldots,j_{k-1}),$$
represents the  2-adic expansion of integers. \qedd
\end{pf}

$R[G]$ is an algebra over $R$ with dimension $2^k$, and a basis is $\{g\mid g\in G\}$. Note that all these basis elements are idempotent in $R[G]$. Elements of $R[G]$ can also be distinguished with functions from $G$ to $R$. In this sense, $R[G]$ is isomorphic to the $R$-representation of $G$. Since $G=\mathbb{F}_2^{\oplus k}$, the $k$-fold direct sum of $\mathbb{F}_2$, we also have
\begin{equation}\label{tensordecomp}
    \mathcal{H}_k(R)\cong R[G]\cong  R[\mathbb{F}_2]^{\otimes k}\cong  \mathcal{H}_1(R)^{\otimes k}.
\end{equation}
Here $\otimes k$ denotes $k$-fold tensor product of an $R$-algebra. This tensor decomposition of $\mathcal{H}_k(R)$ can also be made explicit. Let $\{e_i\mid 0\leq i\leq k-1\}$ be the standard basis of $G$ over $\mathbb{F}_2$, i.e., $e_i$ has $i$-th component 1 and all other components 0 ($0\leq i\leq k-1$). Then any $g\in G\backslash\{e\}$ can be represented as $g=e_{i_1}\oplus e_{i_2}\oplus\cdots\oplus e_{i_s}$ for certain $0\leq i_1< i_2<\cdots< i_s\leq k-1$. It is easy to check that the Hadamard matrix corresponding to this $g$ under the isomorphism in Theorem \ref{isogpring}, which is actually a permutation matrix, can be decomposed into 
\[I_2\otimes\cdots\otimes I_2\otimes \stackrel{i_1}{J_{2}}\otimes I_2\otimes\cdots\otimes I_2\otimes \stackrel{i_2}{J_{2}}\otimes\cdots\otimes \stackrel{i_s}{J_{2}}\otimes\cdots\otimes I_2~({k~\text{terms~in~total}}),\]
where 
$$I_2=\begin{pmatrix}
    1&0\\0&1
\end{pmatrix},~~
J_2=\begin{pmatrix}
    0&1\\1&0
\end{pmatrix}.$$
Besides, the Hadamard matrix corresponding to $e$ is obviously $I_{2^k}=I_2^{\otimes k}$. Note that $I_2$ and $J_2$ form the basis of $\mathcal{H}_1(R)$ and $J_2^2=I_2$. Therefore, under the conversion that $J_2^0=I_2$, the isomorphism \eqref{tensordecomp} implies that any $2^k\times 2^k$ Hadamard matrix over $R$ can be decomposed into a polynomial-like form, that is,
\begin{equation}\label{polyfac}
    A=\sum_{i=0}^{2^k-1}a_iJ_2^i,~~a_i\in R,
\end{equation}
where 
$$J_2^i:=J_2^{i_0}\otimes J_2^{i_1}\otimes\cdots\otimes J_2^{i_{k-1}},~\text{for}~i=\sum_{l=0}^{k-1}i_l2^{k-1-l}.$$
From properties of Kronecker products of matrices, we have 
$$J_2^i\cdot J_2^j= (J_2^{i_0}\cdot J_2^{j_0})\otimes \cdots \otimes (J_2^{i_{k-1}}\cdot J_2^{j_{k-1}}) =J_2^{i\oplus j}.$$ 
Hence this polynomial-like representation for Hadamard matrices indeed induces  an isomorphism between $\mathcal{H}_k(R)$ and a polynomial algebra. In this sense, $\mathcal{H}_k(R)$ is also a Clifford algebra over $R$.
\begin{thm}\label{isopoly}
$$\mathcal{H}_k(R)\cong R[x_1,x_2,\ldots,x_k]/(x_1^2-1,\ldots,x_k^2-1).$$
\end{thm}
\begin{pf}
    Under the decomposition \eqref{polyfac} of a Hadamard matrix $A$, we distinguish it with a multivariate polynomial  $\sum_{i=0}^{2^k-1}a_iX^i$ where $X^i:= x_1^{i_0}x_2^{i_1}\cdots x_{{k}}^{{i_{k-1}}}$ for $i=\sum_{l=0}^{k-1}i_l2^{k-1-l}$. It is direct to check this indeed admits the desired isomorphism. \qedd
\end{pf}

Recall that for any group ring, we can define the augmentation map, that is, 
$$\epsilon:~R[G]\longrightarrow R,~~\sum_{g\in G}a_g g\longmapsto \sum_{g\in G}a_g.$$
The kernel $I$ of $\epsilon$ is called the augmentation ideal of $R[G]$. It is easy to prove that, as a sub-algebra of $R[G]$, $I$ has dimension $2^k-1$ with a basis $\{g-e\mid g\in G\backslash\{e\}\}$. 

In the following, we assume $\char(R)=2$. From Proposition \ref{prop1} we know that, by distinguishing a Hadamard matrix over $R$ with an element in $R[G]$, its image under $\epsilon$ is just the eigenvalue. Therefore, all elements in $I$ are nilpotent. When the ring $R$ has no nilpotent elements, the ideal $I$ is just the nilradical of $R[G]$, i.e., intersection of all prime ideals of  $R[G]$. Specifically, when $R$ is a field, we know that $R[G]$ is an Artinian algebra (finite dimensional algebras over fields are Artinian), so $I$ is simultaneously the nilradical and Jacobson radical of $R[G]$. In fact, $I$ is the unique  maximal ideal of $R[G]$ since it has a dimension $2^k-1$ and thus $R[G]$ is a local ring.

The nilpotency degree of an ideal is defined to be the smallest power that will make it vanish. For the ideal
$I$ we talk about, its nilpotency degree can be determined. 

\begin{thm}\label{nildeg}
    Assume $\char(R)=2$ and $I$ is the augmentation ideal of the group ring $R[G]$. Then the nilpotency degree of $I$ is $k+1$. 
\end{thm}
\begin{pf}
  We prove $I^{k+1}=(0)$ while $I^k\neq (0)$. As $I$ is an $R$-algebra, we need only to prove that any $k+1$ basis elements multiply to 0 while there exist $k$ basis elements that cannot.

  Let $\{e_i\mid 0\leq i\leq k-1\}$ be the standard basis of $G$ over $\mathbb{F}_2$. Then $\{e_i+e\mid 0\leq i\leq k-1\}$ are $k$ basis elements of $I$. Note that
  $$\prod_{i=0}^{k-1}(e_i+e)=\sum_{c_0,c_1,\ldots,c_{k-1}\in \mathbb{F}_2}\bigoplus_{i=0}^{k-1} c_ie_i=\sum_{g\in G} g\neq0$$
in $R[G]$.

On the other hand, let $\{g_i+e\mid g_i\in G,~0\leq i\leq k\}$  be any $k+1$ basis elements of $I$. We can assume they are pairwise distinct, since otherwise they will multiply to 0 naturally. Then
$$\prod_{i=0}^{k}(g_i+e)=\sum_{c_0,c_1,\ldots,c_k\in \mathbb{F}_2}\bigoplus_{i=0}^k c_ig_i.$$
Note that this sum iterates over all  $\mathbb{F}_2$-linear combination of $\{g_i\mid 0\leq i\leq k\}$. As $\{g_i\mid 0\leq i\leq k\}$ must be linearly dependent over $\mathbb{F}_2$, each term turns out to appear $2^r$ times in the sum (in fact, $r=k-\rk_{\mathbb{F}_2}\{g_i\mid 0\leq i\leq k\}$). Therefore, the sum vanishes since $\char(R[G])=2$.\qedd
\end{pf}

Theorem \ref{nildeg} indicates that any $k+1$ elements of $I$ multiply to 0, which coincides with \cite[Proposition 7]{KR21}. Considering Theorem \ref{isopoly}, the proof of Theorem \ref{nildeg} can become more simpler or in some sense obvious. In fact, it is easy to observe that, under the isomorphisms in Theorem \ref{isogpring} and Theorem \ref{isopoly}, the augmentation ideal $I$ of $R[G]$ is isomorphic to the ideal $$I'=(x_1-1,x_2-1,\ldots,x_k-1)$$ of the polynomial algebra $R[x_1,x_2,\ldots,x_k]/(x_1^2-1,\ldots,x_k^2-1)$. Note that $I'^{k+1}$ is generated by
$$(x_{i_1}-1)(x_{i_2}-1)\cdots(x_{i_{k+1}}-1),~1\leq i_1,i_2,\ldots,i_{k+1}\leq k,$$
which are all 0. This is because any $(k+1)$-term multiplication of $k$ elements must contain duplicate terms, killing itself modulo $x_i^2-1$ for some $1\leq i\leq k$. Therefore, after establishing the isomorphisms in Theorem \ref{isogpring} and Theorem \ref{isopoly}, one can obtain a one-sentence proof of \cite[Proposition 7]{KR21}.

% \section{Generalized Hamamard matrices over commutative rings}
% Theorem \ref{isogpring} suggests us to generalize the definition of Hadamard matrices over a commutative ring $R$ by considering  $R[G]$ for a general Abelian group $G$. Here we only discuss the most interested case $G=(\mathbb{F}_q^k,+)$ where $q$ is a prime power.

\section{Block-Hadamard matrices and Hadamard-block matrices}\label{sec:blockhad}

Let $R$  be any unital ring (not necessarily commutative). We can also define Hadamard matrices over $R$ like in \eqref{hij} for a given vector of lenth $2^k$ over $R$. Then it is direct to check the set of all such matrices, also denoted by $\mathcal{H}_k(R)$, forms a ring. Note that $\mathcal{H}_k(R)$ is not commutative if $R$ is not. Specially, if $R$ is the matrix ring, we call a Hadamard matrix over $R$ a block-Hadamard matrix. Indeed, it can be viewed as a block matrix of the Hadamard type. On the other hand, like in Conjecture \ref{conj1} we can also consider block matrices whose blocks are all of Hadamard type, and we call them Hadamard-block matrices. 

Now assume $R$ is a commutative ring. Then the sets of all $2^ks\times 2^ks$ block-Hadamard matrices and Hadamard-block matrices over $R$ admit the algebras $\mathcal{H}_k(\mathscr{M}_{s\times s}(R))$ and $\mathscr{M}_{s\times s}(\mathcal{H}_k(R))$, respectively. Note here that neither of the two algebras is commutative. We have the following theorem.

\begin{thm}\label{thm:bh-hb}
    $$\mathscr{M}_{s\times s}(\mathcal{H}_k(R))\cong
		\mathcal{H}_k(\mathscr{M}_{s\times s}(R)).$$
\end{thm}
\begin{pf}
Let $M$ be an $s\times s$ block matrix whose blocks $M_{u,v}$, $0\leq u,v\leq s-1$, are all $2^k\times 2^k$ Hadamard matrices over $R$.  We call $M_{u,\bullet}$, $0\leq u\leq s-1$, a block row of $M$ and respectively, $M_{\bullet,v}$, $0\leq v\leq s-1$, a block column. Note that $M_{u,\bullet}$ contains $2^k$ rows and $M_{\bullet,v}$ contains $2^k$ columns. 
Now we perform row and column permutations to $M$ to arrange it to be a $2^k\times 2^k$ block matrix $\tilde{M}$ with $s\times s$ blocks. Firstly, move the 0-th row of each block row $M_{u,\bullet}$ to the top of the matrix keeping their ordering, which will form a new block row $M_{0,\bullet}'$ containing $s$ rows; secondly,  move the 1-st row of each block row $M_{u,\bullet}$ to the rows under $M_{0,\bullet}'$ keeping their ordering, which will form a new block row $M_{1,\bullet}'$ containing $s$ rows. Repeat this process to form the new block row $M_{i,\bullet}'$ by moving the $i$-th row of each $M_{u,\bullet}$ for any $2\leq i\leq 2^k-1$. At last we get a matrix $M'$ from $M$. Then, we perform column permutations to block columns $M_{\bullet,v}'$ (containing $2^k$ columns each) of $M'$ in the same manner as we have done for rows of $M$. Finally, we obtain the matrix $\tilde{M}$ which can be viewed as a $2^k\times 2^k$ block matrix  with $s\times s$ blocks. Assume all the row permutations of $M$ is stored in a permutation matrix $P$. Then it is clear that
\[\tilde{M}=P\cdot M\cdot P^\tau=P\cdot M\cdot P^{-1}.\]

To make the structure of $\tilde{M}$ more clear, we assume each Hadamard-type block $M_{u,v}$ of $M$ is determined by its 0-th row $\left( (a_{uv})_0, (a_{uv})_1,\ldots, (a_{uv})_{2^k-1} \right)$, $0\leq u,v\leq s-1$. Then from the above process to derive   $\tilde{M}$, one can figure out that the 0-th block row of $\tilde{M}$ contains $2^k$ blocks of size $s\times s$, say, $A_{0}$, $A_{1}$, $\ldots$, $A_{2^k-1}$, such that
\[
A_{i} = \left( (a_{uv})_i \right)_{u,v=0}^{s-1},~~0\leq i\leq 2^k-1,
\]
and all the other block rows of $\tilde{M}$ are determined by the 0-th block row in the manner that  $\tilde{M}_{i,j}=A_{i\oplus j}$, $0\leq i,j\leq 2^k-1$. This means  $\tilde{M}$ is a block-Hadamard matrix. 
Therefore, the map $M\mapsto\tilde{M}$ can establish the isomorphism between $\mathscr{M}_{s\times s}(\mathcal{H}_k(R))$ and $\mathcal{H}_k(\mathscr{M}_{s\times s}(R))$.
    \qedd
\end{pf}

From Theorem \ref{thm:bh-hb} and the proved  Conjecture \ref{conj1}, we have the following interesting result.

\begin{cor}
    Let $R$ be a commutative ring with characteristic 2, and let $H\in\mathcal{H}_k(\mathscr{M}_{s\times s}(R))$ with first row $(H_0,H_1,\ldots,H_{2^k-1})\in \mathscr{M}_{s\times s}(R)^{2^k}$.  Denote by $q(x)$ the characteristic polynomial of $\sum_{i=0}^{2^k-1}H_i$.
    Then $${q(H)^2 = 0}.$$
\end{cor}
\begin{pf}
Let $P$ be the permutation matrix defined in the proof of Theorem \ref{thm:bh-hb}. Then we know that $H$ corresponds to a matrix $M\in \mathscr{M}_{s\times s}(\mathcal{H}_k(R))$ in the manner that
\[ M = P^{-1}\cdot H\cdot P. \]
From the proof of Theorem \ref{thm:bh-hb} we can see that $(\sum_{i=0}^{2^k-1}H_i)_{u,v}=\lambda(M_{u,v})$, the eigenvalue of $M_{u,v}$ for any $0\leq u,v\leq s-1$. This means $\sum_{i=0}^{2^k-1}H_i$ just equals $M''$ as defined in Conjecture \ref{conj1}, and as a result, $q(x)$ is the characteristic polynomial of $M''$. According to the proved Conjecture \ref{conj1}, we have
\[ q(M)^2 = q(P^{-1} H P)^2=P^{-1}\cdot q(H)^2\cdot P=0, \]
which implies ${q(H)^2 = 0}.$
    \qedd
\end{pf}

\section{Conclusion and further work}\label{sec:conclu}
In this paper, we prove the conjecture posed by Keller and Rosemarin at Eurocrypt 2021 \ref{conj1} and thus give an affirmative answer to their open problem on an improved lower bound for dimension of invariant subspace of the Starkad cipher. 
We further study the  set formed by all Hadamard matrices over commutative rings and reveal its algebraic structure from the perspectives of group algebra and  polynomial algebra. It turns out that these characterizations can help us to understand properties of Hadamard matrices deeper and easier. In particular, the group algebra approach can promise generalizations of Hadamard matrices by considering other Abelian groups $G$ instead of  $(\mathbb{F}_2^k,\oplus)$, which have potential applications in the designs of linear diffusion layers for classical and arithmetization-oriented symmetric ciphers. We will study this topic in a further work.

%-------------------------------------------------------------------------------------------------------------------

\end{document}